\documentclass{IEEEtran}
\normalsize

\ifCLASSINFOpdf
\else
\fi

\usepackage{comment} 

\usepackage{cite}
\usepackage{amsmath,amssymb,amsfonts}
\usepackage{textcomp}
\usepackage{bm}
\usepackage[pdftex]{graphicx}
\usepackage{algorithmic}
\usepackage{algorithm}
\usepackage{multicol}
\usepackage{lipsum, color}
\usepackage{mathtools, cuted, nccmath}
\usepackage{commath}

\usepackage{epstopdf}
\usepackage{subfigure}
\usepackage[short]{optidef}
\usepackage[utf8]{inputenc}
\usepackage{epsfig}
\DeclareGraphicsExtensions{.eps}
\usepackage{multirow} 

\usepackage{subfig}

\usepackage{xcolor, soul}
\sethlcolor{red}

\usepackage [english]{babel}
\usepackage [autostyle, english = american]{csquotes}
\MakeOuterQuote{"}

\begin{document}
\title{{\huge Hybrid Precoding Design for Reconfigurable Intelligent Surface aided mmWave Communication Systems}}
\author{Chandan~Pradhan,~\IEEEmembership{Student Member,~IEEE,}
       Ang~Li,~\IEEEmembership{Member,~IEEE,}
       Lingyang~Song,~\IEEEmembership{Fellow,~IEEE,}
       Branka~Vucetic,~\IEEEmembership{Fellow,~IEEE,}
       and~Yonghui~Li,~\IEEEmembership{Fellow,~IEEE \vspace{-2em}}
    


\thanks{Chandan Pradhan, Ang Li, Yonghui Li and Branka Vucetic are  with  the  Centre  of  Excellence  in  Telecommunications, School of Electrical and Information Engineering, University of Sydney, Sydney, NSW 2006, Australia. (e-mail: \{chandan.pradhan,  ang.li2, yonghui.li, branka.vucetic\}@sydney.edu.au).}
\thanks{Lingyang Song is with Peking University, Beijing 100871, China (email: lingyang.song@pku.edu.cn).}}



\maketitle

\begin{abstract}

In this letter, we focus on the hybrid precoding (HP) design for the reconfigurable intelligent surface (RIS) aided multi-user (MU) millimeter wave (mmWave) communication systems. Specifically, we aim to minimize the mean-squared-error (MSE) between the received symbols and the transmitted symbols by jointly optimizing the analog-digital HP at the base-station (BS) and the phase shifts (PSs) at the RIS, where the non-convex element-wise constant-modulus constraints for the analog precoding and the PSs are tackled by resorting to the gradient-projection (GP) method. We analytically prove the convergence of the proposed algorithm and demonstrate the desirable performance gain for the proposed design through numerical results.

\end{abstract}


\vspace{-4mm}

\section{Introduction}

\IEEEPARstart{M}ILLIMETER-WAVE (mmWave) communication, with the capability of exploiting a large stretch of the underutilized spectrum, is a potential technology to meet the high data rate demand of current and future wireless networks \cite{linMMSE2019, angHP2017}. While mmWave suffers from severe path loss, their small wavelength allows the deployment of a large-scale antenna array to provide the high array gain and improve the link quality. Nevertheless, unlike sub-6 GHz communications where fully-digital (FD) processing is employed, the prohibitive cost and power consumption of the hardware components in mmWave bands make the analog-digital hybrid processing a viable solution to reduce the number of power-hungry radio frequency (RF) chains at the mmWave base-station (BS) \cite{linMMSE2019}.


Although the severe path loss can be compensated by the high array gain, the resulting highly-directional beams coupled with high penetration loss from blockage degrade the link quality for mmWave transmissions. Accordingly, for reliable connectivity, the relay-aided mmWave communication systems were proposed as a favourable solution \cite{relayMMWave2017}. However, the use of relays leads to additional operational cost and power consumption incurred by the deployment of RF chains at the relay units, which meanwhile imposes additional thermal noise on the transmitted signal. 

With the recent advances in electromagnetic (EM) meta-surfaces, the reconfigurable intelligent surfaces (RISs) are foreseen as cost-effective and  energy-efficient substitutes for the relay-based systems \cite{di2019smart}. Specifically, the RISs can control the reflection behaviour of the impinging EM signals through the reflecting elements, which are implemented with low-cost programmable passive devices, e.g., positive-intrinsic-negative (PIN) diodes and phase shifters (PSTs) \cite{di2019smart}. Consequently, the RISs generate a spectrally-efficient wireless propagation environment by constructively combining the signals from the reflecting units, without the deployment of additional RF chains and the imposition of thermal noise. Furthermore, the RISs can readily be fabricated in small size and low weight, which can be coated on the buildings' facade, walls, etc.


Due to the above benefits RISs provide, there have been recent efforts on the designs of the RIS-aided multiple-input-single-output (MISO) communication systems \cite{Zhang2019APB, di2019hybrid}. However, in these works, FD precoding is deployed at the BS, which may not be favourable for mmWave communication systems. To date, there are only a limited number of works that consider the RIS-aided mmWave communications \cite{taha2019enabling, wang2019intelligent}. In \cite{taha2019enabling}, the authors focused on the PSs design at the RIS to facilitate channel estimation with single antenna transceivers. While \cite{wang2019intelligent} studied the RIS-based transmission in mmWave communications, it considers the joint FD precoding-PSs design limited to the single-user (SU) system model, where the FD precoding is approximated with the HP design using the spatially sparse precoding approach. The extension of these works to multi-user (MU) communications with joint HP-PSs design is not straightforward, which further involves mitigating the inter-user interference.



In this paper, we consider a MU-MISO mmWave communication system, which is aided by a RIS implemented through the programmable PSTs. Specifically, we jointly optimize the analog-digital HP at the BS and  the PSs at the RIS such that the mean-squared-error (MSE) between the received symbols and the transmitted symbols is minimized. The resulting non-convex optimization problem is solved via the alternating optimization framework, where the gradient-projection (GP) method is adopted to tackle the element-wise constant-modulus constraints for the analog precoding and the PSs. Furthermore, we analytically prove that the proposed algorithm converges to a Karush–Kuhn–Tucker (KKT) point. Numerical results reveal the desirable performance gain for the proposed joint HP-PSs design.

\textit{Notations}: $y$, ${\bf y}$ and $\mathbf{Y}$ denote scalar, vector and matrix, respectively; $Y_{i,j}$ is the $\left(i,j\right)$-th element  of $\mathbf{Y}$; Transpose and conjugate transpose operators are represented by $\mathbf{Y}^T$ and $\mathbf{Y}^H$, respectively; $\mathbf{\norm{Y}}_F$ and $\norm{\bf y}_2$ denote the Frobenius and $\ell_2$ norm, respectively; $\mathcal{D} \left({\bf y}\right)$ operator diagonalizes ${\bf y}$; $\mathcal{D}_C \left({\bf Y}\right)$ operator extracts the diagonal elements of ${\bf Y}$ into a column vector; $\mathcal{D}_B \left\{{\bf y}_1, \cdots, {\bf y}_n \right\}$ denotes a block diagonal matrix  with ${\bf y}_i$ on the  block-diagonal; ${\rm Tr}\left\{{\bf Y}\right\}$ denotes the trace operator;  Expectation of a random variable is noted by $\mathbb{E}[\cdot]$; ${\Re}$ denotes the real part of a complex number; $|\cdot|$ and $\angle$ returns the absolute value and the argument of a complex number, respectively; $\mathbf{I}_{m} \in \mathbb{R}^{m \times m}$ is the identity matrix. 


\begin{figure}
\centering
\includegraphics[width=3.0in, height=1.5in]{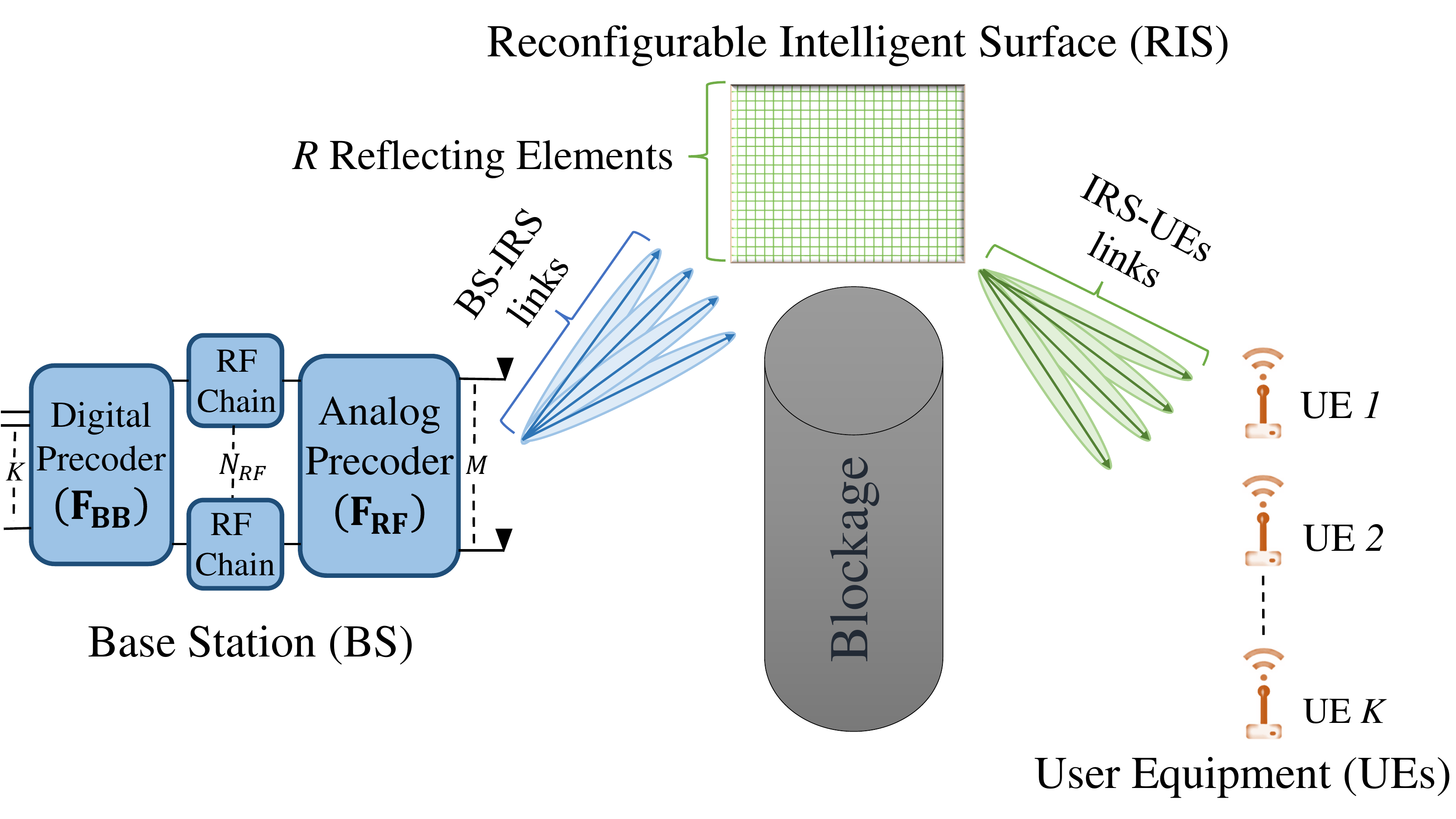}
\vspace{-0.5em}
  \caption{\small{RIS aided MU mmWave system.}}
\vspace*{-5mm}
\end{figure}
\vspace{-4mm}

\section{System Model}

 We consider a RIS aided MU mmWave communication system in the downlink, as shown in Fig.1, where a BS equipped with a uniform linear array (ULA) of $M$ antennas and $N_{RF}$ RF chains communicates with $K$ single-antenna user equipments (UEs) through a RIS with $R$ reflecting elements arranged in the form of a uniform planar array (UPA) \cite{wang2019intelligent}, where $ K \leq N_{RF} \ll M$. During transmission, the BS applies a digital precoder $\mathbf{F_{BB}} \triangleq \left[{\bf f}_1^{\bf BB}, \cdots,{\bf f}_K^{\bf BB} \right] \in \mathbb{C}^{N_{RF} \times K}$  followed by an  analog precoder $\mathbf{F_{RF}} \triangleq \left[{\bf f}_1^{\bf RF}, \cdots, {\bf f}_{N_{RF}}^{\bf RF}\right] \in \mathbb{C}^{M \times N_{RF}}$, where each RF chain is connected to all antennas through the PSTs. Since $\mathbf{F_{RF}}$ is implemented with the PSTs, each entry of $\mathbf{F_{RF}}$ satisfies the element-wise constant-modulus constraint \cite{linMMSE2019}, i.e., ${\bf F_{RF}} \in \mathcal{F}$, where $\mathcal{F} \triangleq \left\{{\bf F_{RF}}  \bigg|  \left|\left[\mathbf{F_{RF}} \right]_{m,n}\right| = \sqrt{\frac{1}{M}}, \forall m,n \right\}$. Assuming that the direct path between the BS and the UEs is blocked and each UE has an adaptive gain control unit at its input as in \cite{MMSELOve2008}, the received signal for the $K$ UEs, after the reflection from the RIS, is obtained as




    \vspace{-2mm}

\begin{equation}
\hat{\bf s} = {\zeta} \mathbf{H}_{I} {\bm \Psi}\mathbf{H}_{B} \mathbf{F_{RF}} {\bf F}_{\bf BB} {\bf s} +  {\zeta}   {\bf n},
\end{equation}
where $\zeta \in \mathbb{R}_{+}$ is the gain induced by the adaptive controller. $\mathbf{s} \in \mathbb{C}^{K \times 1}$ is the transmitted symbol vector which satisfies $\mathbb{E}\left[{\bf s} {\bf s}^H \right] = \frac{P}{K}{\bf I}_K$, $P$ is  the total  transmit power and ${\bf n} \sim \mathcal{C N}(0, \sigma^2 {\bf I}_{K})$ is the additive Gaussian noise vector.  ${\bm \Psi} \triangleq \mathcal{D} \left(\left\{\Psi_1, \cdots, \Psi_R\right\}\right) \in \mathbb{C}^{R \times R}$ is the PS matrix of the RIS, where each $\Psi_r$ satisfies  $\Psi_r \in \mathcal{R} \triangleq \left\{\Psi_r  \bigg| \left|\Psi_r\right| = \sqrt{\frac{1}{R}}, \; \forall r \right\}$. ${\bf H}_{B} \in \mathbb{C}^{R \times M}$ is the mmWave channel matrix between the RIS and the BS, and ${\bf H}_{I} \triangleq \left[{\bf h}_{I_1}, \cdots, {\bf h}_{I_K} \right]^H \in \mathbb{C}^{K \times R}$, where $\mathbf{h}_{I_k}$ is the mmWave channel vector between the $k$-th UE and the RIS. Since the mmWave channels are expected to experience limited scattering, we adopt a geometric channel model with $L_{B}$ and $L_{I}$ propagation paths for the BS-RIS link and the RIS-UEs links, respectively \cite{linMMSE2019, wang2019intelligent}, given by

\begin{subequations}
\begin{equation}
    {\bf H}_{B} = \sqrt{\frac{M R}{L_{B}}} \sum_{l = 1}^{L_{B}} {\xi}_{l} {\bm \alpha}_{P} \left( \psi_{l}^{\left(AoA \right)}, \vartheta_l^{\left(AoA \right)} \right) {\bm \alpha}_{L} \left(\psi_{l}^{\left(AoD \right)} \right)^H, 
\end{equation}
\begin{equation}
  \mathbf{h}_{I_k} = \sqrt{\frac{R}{L_{I}}} \sum_{l = 1}^{L_{I}} {\gamma}_{k,l} {\bm \alpha}_{P} \left(\theta_{k,l}^{\left(AoD \right)}, \phi_{l}^{\left(AoD \right)} \right), \; \forall k, 
\end{equation}
\end{subequations}
where ${\xi}_{l}$ is the complex gain following $\mathcal{CN}\left(0,\sigma^2_{\xi}\right)$, and  $\psi_{l}^{(AoA)} \left(\vartheta_{l}^{(AoA)}\right)$ and $\psi_{l}^{(AoD)}$ are the respective azimuth (elevation) angle-of-arrival (AoA) and angle-of-departure (AoD) of the $l$-th propagation path for the BS-RIS link. ${\bm \alpha}_L \left(\psi \right) \triangleq \frac{1}{\sqrt{N}} \left[1, \cdots, e^{\jmath \frac{2 \pi}{\lambda} d n  \sin(\psi)}, \cdots \right]^T$ and ${\bm \alpha}_P \left(\theta, \phi \right) \triangleq \frac{1}{\sqrt{N}} \left[1, \cdots, e^{\jmath \frac{2 \pi}{\lambda} d \left(p \sin(\theta) \sin(\phi) + q cos(\phi)\right)}, \cdots \right]^T$, where $0 \leq n \leq \left(N-1\right)$, $0 \leq \left\{p,q\right\} \leq \left(\sqrt{N}-1\right)$, $d$ is the antenna spacing and $\lambda$ is the carrier wavelength. The definitions for parameters in ${\bf h}_{I_k}$ are similar and therefore omitted for brevity.

\vspace{-2mm}

\section{Problem Formulation and Proposed Solution}

In this work, we aim to jointly design the HP at BS and PSs at the RIS to minimize the MSE, given by ${\rm MSE} = \mathbb{E} \left[\norm{{\bf s} - \hat{\bf s}}^2 \right]$, which is further expanded into 

\vspace{-1mm}

\begin{equation}
 {\rm MSE} =  P - \frac{2P}{K}{\zeta} {\Re} \left({\rm Tr} \left\{{\bf F}^H {\bf H}^H \right\} \right)+ \frac{P{\zeta^2}}{K}\norm{{\bf H} {\bf F}}^2_F + \sigma^2{\zeta^2}K,
\end{equation}
where ${\bf H} \triangleq {\bf H}_{I} {\bm \Psi} {\bf H}_{B}$ and ${\bf F} \triangleq {\bf F}_{\bf RF} {\bf F}_{\bf BB}$. Accordingly, we formulate the following optimization problem:

\vspace{-1mm}
\begin{equation}
\begin{aligned}
& \mathcal{P}_1:&& \underset{\left\{{\bf F}_{\bf RF}, {\bf F}_{\bf BB}, {\bm \Psi}, \zeta \right\}} \min  {\rm MSE}\\
& \text{\it s.t.}
& &  {\rm C}_1: {\bf F_{RF}} \in \mathcal{F}, \; {\rm C}_2: {\Psi}_r \in \mathcal{R}, \; \forall r, \\
&&& {\rm C}_3: \norm{{\bf F}_{\bf RF} {\bf F}_{\bf BB}}_F^2 = K, \; {\rm C}_4: \zeta \in \mathbb{R}_{+}, \\
 \end{aligned}
\end{equation}
where the power constraint ${\rm C}_3$ of the precoder assumes total transmit power is used at the BS. However, along with the non-convex element-wise constant-modulus constraints ${\rm C}_1$ and ${\rm C}_2$, the coupling between ${\bf F}_{\bf RF}$ and  ${\bf F}_{\bf BB}$ in ${\rm C}_3$ makes $\mathcal{P}_1$ difficult to solve. Subsequently, following \cite{linMMSE2019, MMSELOve2008}, we substitute $\bar{\bf F}_{\bf BB}$ for $\zeta {\bf F}_{\bf BB}$, such that minimizing (3) is equivalent to minimizing 
\vspace{-1mm}
\begin{equation}
 \overline{\rm MSE} =   P - \frac{2P}{K}{\Re} \left({\rm Tr} \left\{\bar{\bf F}^H {\bf H}^H \right\} \right) +\frac{P}{K} \norm{{\bf H} \bar{\bf F}}^2_F +  \sigma^2 \norm{\bar{\bf F}}^2_F,
\end{equation}
where $\bar{\bf F} \triangleq {\bf F}_{\bf RF} \bar{\bf F}_{\bf BB}$, which follows $\norm{\bar{\bf F}}_F^2 = \zeta^2 K$. Consequently, $\zeta$ satisfying ${\rm C}_3$ and ${\rm C}_4$ is given by  $\zeta = \sqrt{\frac{1}{K}}\norm{\bar{\bf F}}_F$, which leads to the following reformulation for $\mathcal{P}_1$: 

\vspace{-1mm}

\begin{equation}
\begin{aligned}
& \mathcal{P}_2: \underset{\left\{{\bf F}_{\bf RF}, \bar{\bf F}_{\bf BB}, {\bm \Psi} \right\}} \min \rm{\overline{MSE}} \; \; \text{\it s.t.} \; {\rm C}_1, {\rm C}_2, \\
 \end{aligned}
\end{equation}
which is still non-convex. To find a feasible solution, we propose to solve $\mathcal{P}_2$ with an iterative method by alternately optimizing the variables $\left\{{\bf F}_{\bf RF}, \bar{\bf F}_{\bf BB}, {\bm \Psi} \right\}$, as detailed in the following sub-sections.


%


\vspace{-2mm}

\subsection{Design of $\bar{\bf F}_{\bf BB}$}

At the $t$-th iteration, $\bar{\bf F}_{\bf BB}^{(t)}$ for a given $\left\{{\bf F}_{\bf RF}^{({t-1})}, {\bm \Psi}^{({t-1})} \right\}$ is obtained from the following sub-problem:

\vspace{-1mm}

\begin{equation}
 \mathcal{P}_3: \underset{\left\{\bar{\bf F}_{\bf BB}^{(t)} \right\}} \min  \overline{\rm MSE}, 
\end{equation}
%
%
which can be solved by taking the gradient of (5) with respect to (w.r.t.) $\bar{\bf F}_{\bf BB}^{(t)}$ and equating it to zero, thereby obtaining

\begin{equation}
 \bar{\bf F}_{\bf BB}^{(t)} = \left[\left({\bf F}_{\bf RF}^{({t-1})}\right)^H {\bm \Xi}^{({t-1})} {\bf F}_{\bf RF}^{({t-1})} \right]^{-1} \left({\bf H}^{({t-1})} {\bf F}_{\bf RF}^{({t-1})}\right)^H ,
\end{equation}
where ${\bm \Xi}^{({t-1})} \triangleq \left[\left({\bf H}^{({t-1})}\right)^H {\bf H}^{({t-1})} + \frac{K\sigma^2}{P} {\bf I}_M \right]$.
%

 %
 
 \vspace{-2mm}

\subsection{Design of ${\bf F}_{\bf RF}$}

Next, with the obtained ${\bf F}_{\bf BB}^{(t)}$ and a given $\left\{{\bf F}_{\bf RF}^{({t-1})}, {\bm \Psi}^{({t-1})} \right\}$, ${\bf F}_{\bf RF}^{({t})}$ is updated by solving the following sub-problem:

\vspace{-2mm}

\begin{equation}
\begin{aligned}
& \mathcal{P}_4: \underset{\left\{{\bf F}_{\bf RF}^{(t)} \right\}} \min  \overline{\rm MSE} \; \; \text{\it s.t.} \;  {\rm C}_1,
 \end{aligned}
\end{equation}
which is a non-convex problem due to ${\rm C}_1$ and is solved using the gradient projection (GP) method. Specifically, the GP method descends towards the optimal solution by projecting each subsequent point at the $t$-th iteration onto the feasible region ${\rm C}_1$, given by

\vspace{-2mm}

\begin{equation}
    {\bf F}_{\bf RF}^{(t)} = \sqrt{\frac{1}{M}} e^{\jmath \angle \left( {\bf F}_{\bf RF}^{({t-1})} + \alpha_{B} \mathbf{d}^{({t-1})}\right)},
\end{equation}
%
%
%
%
while moving along the descent direction ${\bf d}^{({t-1})}  \triangleq - {\nabla {\bf g} \left({\bf F}_{\bf RF}^{({t-1})} \right)}$ with a step size $\alpha_{B}$, where the Euclidean gradient of the unconstrained objective of  $\mathcal{P}_{4}$ w.r.t. ${\bf F}_{\bf RF}^{(t-1)}$, i.e., $\nabla {\bf g} \left({\bf F}_{\bf RF}^{(t-1)} \right)$, is given by

\vspace{-2mm}

\begin{equation}
\nabla {\bf g} \hspace{-0.25em} \left({\bf F}_{\bf RF}^{(t-1)} \right) = \frac{P}{K} \left[{\bm \Xi}^{({t-1})}{\bf F}_{\bf RF}^{(t-1)} \bar{\bf F}_{\bf BB}^{(t)} - \left({\bf H}^{({t-1})}\right)^H \right]\hspace{-0.35em}\left(\bar{\bf F}_{\bf BB}^{(t)}\right)^H.
\end{equation}

\textit{\textbf{Lemma 1:}} The cost function of $\mathcal{P}_{4}$ decreases at each iteration if the step-size satisfies that $\alpha_{B} \leq \frac{1}{\tau^{(t)}}$, where $\tau^{(t)} \triangleq \left(\frac{P}{K}\norm{{\bf H}^{({t-1})}}_F^2 + \sigma^2 \right) \norm{\bar{\bf F}_{\bf BB}^{(t)}}^2_F$.

\textit{\textbf{Proof:}}  Refer to Appendix \ref{FirstAppendix}

{\textit{Remark:} As detailed in the Appendix \ref{SecondAppendix}, the proposed analog precoding design can be readily extended to the scenario when partially-connected structure (PCS) is employed at the BS, where the array gain is sacrificed to further improve the energy-efficiency by connecting each RF chain to  a subset of $\frac{M}{N_{RF}}$ antennas, in which case only a total number of $M$ PSTs are required \cite{angHP2017}. 

\vspace{-4mm}

\subsection{Design of ${\bm \Psi}$}

Finally, for the obtained $\left\{\bar{\bf F}_{\bf BB}^{(t)}, {\bf F}_{\bf RF}^{(t)}\right\}$ and a given ${\bm \Psi}^{({t-1})}$, ${\bm  \Psi}^{(t)}$ is updated by solving the following sub-problem:

\begin{equation}
\begin{aligned}
& \mathcal{P}_5: \underset{\left\{{\bm \Psi}^{(t)} \right\}} \min  \; \overline{\rm MSE} \; \; \text{\it s.t.} \; {\rm C}_2, 
 \end{aligned}
\end{equation}
where ${\rm C}_2$ makes the problem non-convex. Accordingly, $\mathcal{P}_5$ is solved using the GP method similar to $\mathcal{P}_4$ with appropriate modifications owing to the diagonal structure of ${\bf \Psi}^{(t)}$. Specifically, ${\bm \Psi}^{(t)}$ is updated by projecting each subsequent point at the $t$-th iteration onto the feasible region ${\rm C}_2$, given by

\begin{equation}
    {\bm \Psi}^{(t)}  = \sqrt{\frac{1}{R}}\mathcal{D}\left(e^{\jmath \angle \left(\mathcal{D}_C \left({\bm \Psi}^{({t-1})} + \alpha_{I} {\bm \delta}^{({t-1})}\right)\right)}\right),
\end{equation}
%
%
%
while moving along the descent direction ${\bm \delta}^{({t-1})} \triangleq  - {\nabla {\bf g} \left({\bm \Psi}^{({t-1})} \right)}$ with a step size $\alpha_{I}$, where the Euclidean gradient of the unconstrained objective of  $\mathcal{P}_{5}$ w.r.t. ${\bm \Psi}^{(t-1)}$, i.e., $\nabla {\bf g} \left({\bm \Psi}^{(t-1)} \right)$, is given by 
%
\vspace{-2mm}

\begin{equation}
\begin{aligned}
  &\nabla {\bf g} \left({\bm \Psi}^{(t-1)} \right) = \mathcal{D} \left( \mathcal{D}_C \left(  \ddot{\bm \delta}^{({t-1})} \right)\right),
  \end{aligned}
\end{equation}
%
%
where $\ddot{\bm \delta}^{({t-1})} \triangleq \frac{P}{K}{\bf H}_{I}^H \left({\bf H}_{I} {\bm \Psi}^{({t-1})} \bar{\bm \Gamma}^{(t)} - {\bf I}_{K} \right) \left(\bar{\bm \Gamma}^{(t)}\right)^H$ and $\bar{\bm \Gamma}^{(t)} \triangleq {\bf H}_{B} {\bf F}_{\bf RF}^{(t)} \bar{\bf F}_{\bf BB}^{(t)}$. 

\textit{\textbf{Lemma 2:}} The cost function of $\mathcal{P}_{5}$ decreases at each iteration if the step-size satisfies that $\alpha_{I} \leq \frac{1}{\varsigma^{(t)}}$, where $\varsigma^{(t)} \triangleq \frac{P}{K} \norm{\bar{\bm \Gamma}^{(t)}}_F^2 \norm{{\bf H}_{I}}_F^2$.

{\textit{\textbf{Proof:}} The proof is similar to that of Lemma 1, which is therefore omitted for brevity. $\blacksquare$

Algorithm 1 summarizes the proposed framework for the joint HP-PSs design for the RIS-aided mmWave system.


\vspace{-2mm}

\begin{algorithm}[!htb]
 \begin{algorithmic}[1]
\STATE \textbf{Input}: ${\bf H}_{I}$, ${\bf H}_{B}$.
\STATE Initialize  ${\bf F}_{\bf RF}^{(0)}$ and ${\bm \Psi }^{(0)}$ with random phase, $t \gets 1$; 
\REPEAT
\STATE  Update $\bar{\bf F}^{(t)}_{\bf BB}$ using (8);
\STATE  Update ${\bf F}_{\bf RF}^{(t)}$ using (10);
\STATE Update ${\bm \Psi}^{(t)}$ using (13);
\STATE $t \gets t + 1$;
\UNTIL{convergence}
\STATE Normalize ${\bf F}_{\bf BB} = \frac{1}{\zeta} \bar{\bf F}_{\bf BB}$;
\STATE \textbf{Output}: ${\bf F}_{\bf RF}$ and ${\bf F}_{\bf BB}$.
\end{algorithmic}
\caption{ Proposed joint HP and PSs design}
\label{GP_AG}
\vspace{-0.35em}
\end{algorithm}


\textit{\textbf{Lemma 3:}} The cost function of $\mathcal{P}_{2}$ decreases at each iteration of Algorithm 1 and the solution converges to a KKT point.

\vspace{-1mm}

\textit{\textbf{Proof:}}  Given Lemma 1 and Lemma 2, the cost function of $\mathcal{P}_{2}$ is positive and minimized within each iteration at Step 4, Step 5 and Step 6, which accordingly converges to a locally optimal solution. Upon convergence, $\bar{\bf F}_{\bf BB}$ given by (8) for the obtained $\left\{{\bf F}_{\bf RF}, {\bm \Psi} \right\}$ is seen to satisfy the KKT points of $\mathcal{P}_{2}$. Next, observing $\nabla h\left({\bf F}_{\bf RF}^{({t-1})}\right) = \nabla \tilde{h}\left({\bf F}_{\bf RF}^{({t-1})}; {\bf F}_{\bf RF}^{({t-1})}\right)$, and since the solution sequence $\left\{{\bf F}_{\bf RF}^{(t)}\right\}_t$ lies in a compact set, according to  Bolzano–Weierstrass Theorem and \cite[Theorem 1]{Razaviyayn2013}, ${\bf F}_{\bf RF}$ given by (10) for the obtained $\left\{\bar{\bf F}_{\bf BB}, {\bm \Psi} \right\}$ converges to a KKT point of $\mathcal{P}_{2}$. Similar explanation holds for ${\bm \Psi}$. $\blacksquare$



%

\vspace{-4mm}

\subsection{Complexity analysis}

We analyze the complexity of the proposed HP-PSs design per iteration, considering only the dominant components of the calculations. In this work, we assume that the number of active RF chains is equal to the number of UEs, i.e. $N_{RF} = K$. Accordingly, it  can  be  seen  that  the  complexity  of the proposed design given in Algorithm 1 is mainly incurred from the calculation of a) ${\bf F}_{\bf BB}$, b) $\nabla {\bf g} \left({\bf F}_{\bf RF} \right)$, and c) $\nabla {\bf g} \left({\bm \Psi} \right)$, whose corresponding computational costs are $\mathcal{O}\left(2 M^2 K \right)$, $\mathcal{O}\left(2 M^2 K\right)$, and $\mathcal{O} \left(3 M R^2 + R^3 \right)$, respectively. Hence, the overall complexity of the proposed algorithm is $\mathcal{O}\left( 4 M^2 K + 3 M R^2 + R^3\right)$.

\vspace{-4mm}

\section{Numerical Results}

In  this  section,  we  evaluate  the  performance of our proposed design via Monte-Carlo simulations. Unless stated otherwise, we assume $M = 48$, $N_{RF} = K = 6$, $R = 100$, $L_B = L_I = 5$ for the mmWave channels, and SNR is defined as $ {\rm SNR} \triangleq 10 \; \log_{10} \frac{1}{K \sigma^2}$, where the total transmit power is set as $P = 1$. The  antenna spacing is $d = \frac{\lambda}{2}$. To evaluate the performance of the proposed joint HP-PSs design, we use the following baseline schemes: 1) \textit{upBound:} FD precoding at the BS with a FD RIS (upper bound), 2) \textit{fdBS-optRIS:} FD precoding at the BS with optimized phase for RIS, 3) \textit{hpBS-rndRIS:} HP at the BS with random phase for RIS, and 4) \textit{hpBS-noRIS:} HP at the BS with no RIS in the considered system. Subsequently, we evaluate the proposed design in terms of the spectral efficiency, defined similar to that in \cite{angHP2017}.


\vspace{-4mm}
\begin{figure}[!htb]
 \centering
   \subfigure[ ]
    {
        \includegraphics[scale=0.1625]{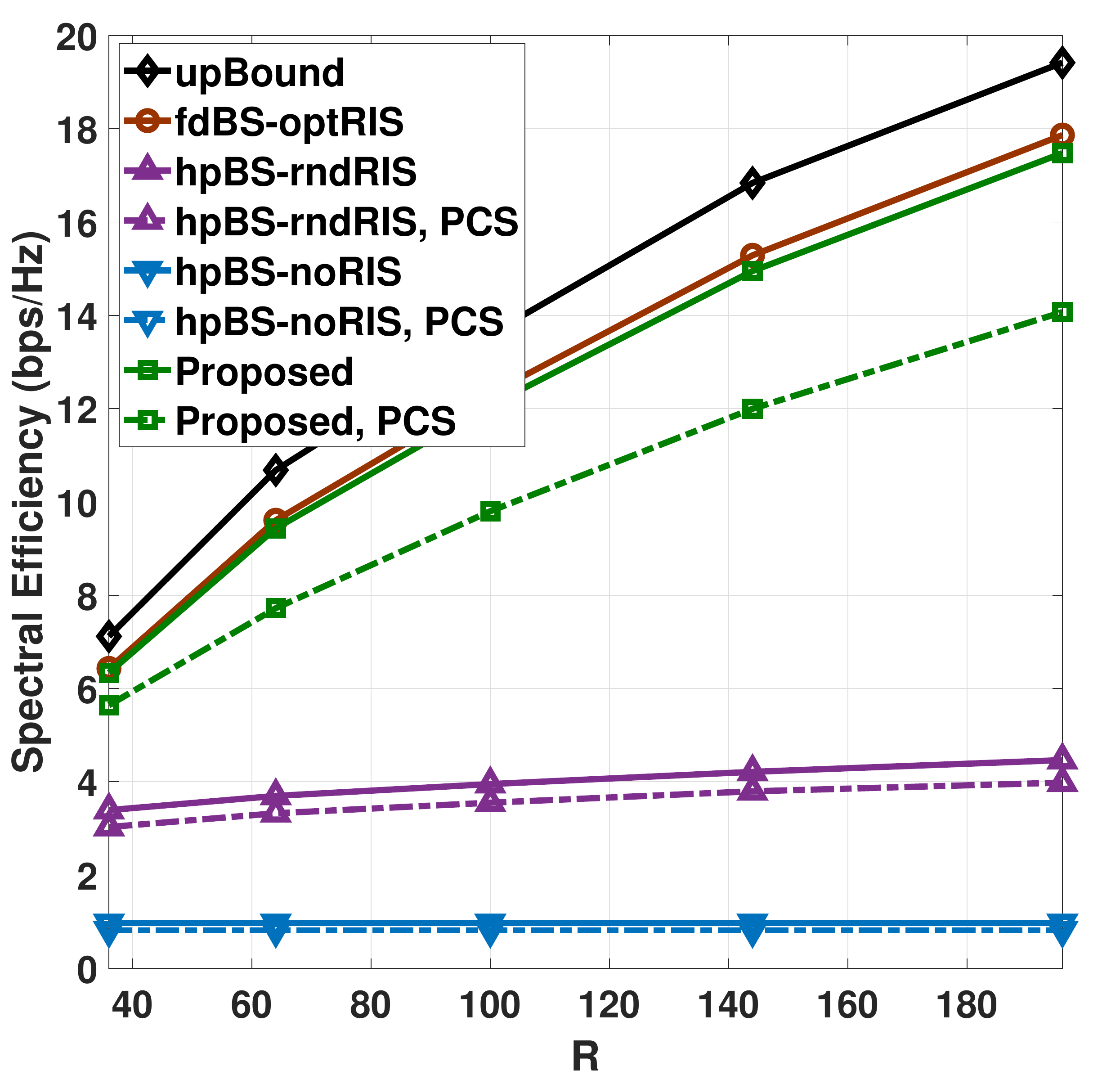}
    }  \hskip -2.0ex
    \subfigure[] 
    {
         \includegraphics[scale=0.1635]{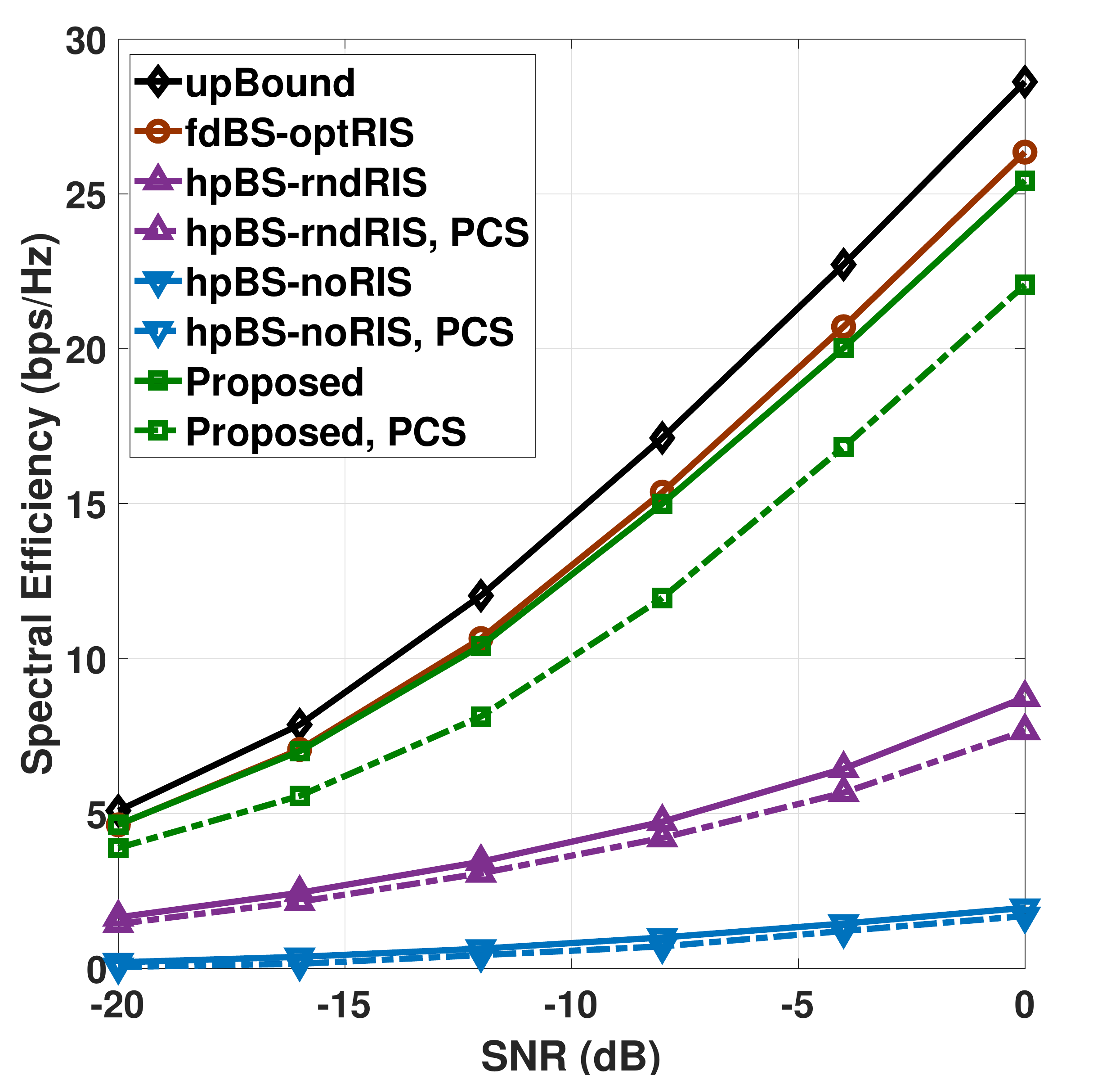}
 }     \vspace{-3mm}
    \caption{\small Spectral efficiency v.s. a) $R$, $M = 48$, $N_{RF} = K = 6$, ${\rm SNR} = -10 \; {\rm dB}$, b) ${\rm SNR}$, $M = 48$, $N_{RF} = K = 6$, $R = 100$.}
    \vspace{-2mm}
\end{figure}

 The SE performance of the proposed design w.r.t $R$ at ${\rm SNR} = -10 {\rm \; dB}$ is shown in Fig.2(a), where SE improves with an increase in $R$. Though the proposed design suffers a performance loss compared to \textit{upBound} and \textit{fdBS-optRIS}, owing to the deployment of HP at the BS and constant-modulus PSs at the RIS, it enjoys a significantly higher energy-efficiency requiring only $6$ RF chains at the BS. Furthermore, the proposed design achieves a noticeable gain over \textit{hpBS-rndRIS} and \textit{hpBS-noRIS}. Note that the limited array gain leads to the SE loss when PCS is employed at the BS. In Fig.2(b), a similar  trend  can  be  observed  for  the proposed design w.r.t ${\rm SNR}$ for $R = 100$, where the SE increase w.r.t ${\rm SNR}$.
\vspace{-3mm}

\section{Conclusion}

In this paper, we have proposed an alternating optimization framework to jointly design the HP and the PSs for the RIS aided MU mmWave communication systems. Specifically, an iterative algorithm was devised to minimize the MSE by jointly optimizing the HP at the BS and the PSs at the RIS, where the GP method was leveraged to tackle the element-wise constant-modulus constraints for the analog precoding and the PSs. The convergence of the proposed algorithm is analytically proved and the desired performance gain of the proposed design has been validated through numerical examples.

\vspace{-3mm}

\appendix

  \subsection{Proof for Lemma 1}
  \label{FirstAppendix}
  
 Given the solution of $\mathcal{P}_{4}$ at the $\left(t-1\right)$-th iteration, i.e., ${\bf F}_{\bf RF}^{({t-1})}$ and by defining $\Delta {\bf F}_{\bf RF} \triangleq \left({\bf F}_{\bf RF}^{(t)} - {\bf F}_{\bf RF}^{({t-1})}\right)$, the upper bound for the cost function of  $\mathcal{P}_{4}$  at the $t$-th iteration, denoted by $h\left({\bf F}_{\bf RF}^{(t)}\right)$, can be written as 

\vspace{-2mm}
\begin{equation}
\begin{aligned}
h\left({\bf F}_{\bf RF}^{(t)}\right)  &\leq  \tilde{h}\left({\bf F}_{\bf RF}^{(t)}; {\bf F}_{\bf RF}^{({t-1})} \right) = h\left({\bf F}_{\bf RF}^{({t-1})}\right) \\
& \hspace{1.5em}-  2{\Re} \left( {\rm Tr}\left\{ {\bf d}^{({t-1})} \Delta {\bf F}_{\bf RF}^H  \right\} \right) + {\tau^{(t)}}\norm{\Delta {\bf F}_{\bf RF}}_F^2,
\end{aligned}
\end{equation}
where  $\tau^{(t)} \triangleq \left(\frac{P}{K}\norm{{\bf H}^{({t-1})}}^2_F + \sigma^2 \right) \norm{\bar{\bf F}_{\bf BB}^{(t)}}^2_F$. Subsequently, observing that $h\left({\bf F}_{\bf RF}^{(t)}\right)  \leq \tilde{h}\left({\bf F}_{\bf RF}^{(t)}; {\bf F}_{\bf RF}^{({t-1})}\right), \; \forall t$ and $ h\left({\bf F}_{\bf RF}^{({t-1})}\right) = \tilde{h}\left({\bf F}_{\bf RF}^{({t-1})}; {\bf F}_{\bf RF}^{({t-1})}\right)$, we formulate the following upper bound minimization problem corresponding to $\mathcal{P}_{4}$:

\vspace{-2mm} 

\begin{equation}
\begin{aligned}
& \mathcal{P}_{6}: \underset{\left\{{\bf F}_{\bf RF}^{(t)}\right\}}{\min} \; \tilde{h}\left({\bf F}_{\bf RF}^{(t)}; {\bf F}_{\bf RF}^{({t-1})}\right) \; \text{\it s.t.} \; {\rm C}_1, 
\end{aligned}
\end{equation}
which has the solution given by (10). Accordingly, we have

\vspace{-2mm} 

\begin{equation}
\begin{aligned}
h\left({\bf F}_{\bf RF}^{(t)}\right) &\leq \tilde{h}\left({\bf F}_{\bf RF}^{(t)}; {\bf F}_{\bf RF}^{({t-1})} \right), \\
&\leq \tilde{h}\left({\bf F}_{\bf RF}^{({t-1})}; {\bf F}_{\bf RF}^{({t-1})} \right) = h\left({\bf F}_{\bf RF}^{({t-1})}\right),
\end{aligned}
\end{equation}
which completes the proof. $\blacksquare$

\vspace{-4mm}

  \subsection{${\bf F}_{\bf RF}$ design for PCS}
  \label{SecondAppendix}
  
When the PCS is employed at the BS, the analog precoder, defined by $\tilde{\bf F}_{\bf RF}  \triangleq  \mathcal{D}_B\left(\left[\tilde{\bf f}_1^{\bf RF}, \cdots, \tilde{\bf f}_{N_{RF}}^{\bf RF}\right]\right)  \in \mathbb{C}^{M \times N_{RF}}$, attains a  block diagonal structure and each non-zero element in $\tilde{\bf F}_{\bf RF}$ satisfies $\left|\left[\tilde{\bf F}_{\bf RF} \right]_{m \in \mathcal{I}(n),n}\right|  = \sqrt{\frac{N_{RF}}{M}}, \forall n$, where $\mathcal{I}(n) \triangleq \left[\left(\frac{M}{N_{RF}}\left(n - 1\right) + 1\right), \cdots, \left(\frac{M}{N_{RF}} n \right)\right]$ \cite{angHP2017}. Accordingly, $\tilde{\bf F}_{\bf RF}^{(t)}$ at the $t$-th iteration is updated as 

\begin{equation}
    \tilde{\bf F}_{\bf RF}^{(t)} = \sqrt{\frac{N_{RF}}{M}}
    \mathcal{D}_B\left(\left[e^{\jmath \angle \left(\left[ \tilde{\bf F}_{\bf RF}^{({t-1})} + \alpha_{B} \mathbf{d}^{({t-1})}\right]_{\left(\mathcal{I}\left(n\right),n\right)}\right)}\right]_{\forall n}\right),
\end{equation}
where \textbf{\textit{Lemma 1}} holds for $\tilde{\bf F}_{\bf RF}^{(t)}$.

\ifCLASSOPTIONcaptionsoff
  \newpage
\fi

\vspace{-2mm}

\bibliography{myBib.bib}
\bibliographystyle{IEEEtran}

\end{document}